# Digital Homunculi and Institutional Design: Breaking Through the Experimentation Bottleneck

**Petr Špecián**

petr.specian@fsv.cuni.cz

*Institute of Political Studies, Faculty of Social Sciences, Charles University*

**Abstract:** Democracy research faces a longstanding experimentation bottleneck. Potential institutional innovations remain untested because human-subject studies are slow, expensive, and ethically fraught. This paper argues that *digital homunculi*, that is, GenAI-powered agents role-playing humans in diverse institutional settings, could offer a way to break through the bottleneck. In contrast to the legacy agent-based modeling, building complexity from transparent simple rules, the *digital homunculi* methodology aims to extract latent human behavioral knowledge from opaque large language models. To this ends, it designs multi-agent interactions as elicitation devices to trigger in LLMs human-like behavior that can be recorded as synthetic data. However, the validity of synthetic data remains an open question. Success requires that accurate, coherent, transferable models of humans ("little humans" - homunculi) already lurk within GenAI's inscrutable matrices and can be lured out via the social simulation role-play exercise. At the same time, to the extent these attempts are successful, they promise to completely transform the political economy of institutional research from scarcity to abundance. To help mitigate the number of challenges along the way to such success, I propose concrete validation strategies including behavioral back-testing via knowledge cutoffs, and outline infrastructure requirements for rigorous evaluation. The stakes are high: legacy democratic institutions develop at much slower pace than the surrounding technological landscape. If they falter, we lack a repository of tested backup alternatives. Breaking through the experimentation bottleneck must be a priority and digital homunculi may be quickly maturing into a methodology capable of achieving this feat.

**Keywords:** Artificial Intelligence; Generative Agents; Democratic Innovation; Institutional Design; Synthetic Data; Experimental Political Science

## 1. Introduction

Technology races forward, introducing radically new paradigms in a swift order. The foundation of the modern generative artificial intelligence (GenAI) was invented in 2017 (Vaswani et al., 2017). By late 2025, GenAI is already reshaping the world. Meanwhile, key democratic infrastructure remains largely static. Even minor institutional tweaks are divisive and rarely get deployed. The 'pacing problem' (Marchant 2011) is real and dangerous.

Since democracy is a system where free debate among equals plays a key part, no significant disruption in the ways knowledge is produced, disseminated, and assimilated is to be taken lightly (cf. Kurtulmus 2020). Digitalization of democratic discourse with its epistemic democratization, viral falsehoods, and echo chambers has been posing significant and yet unresolved challenges (e.g., Sunstein 2017; Gurri 2018, Špecián 2022a). Nowadays, advanced AI systems are swiftly arriving on the scene, threatening 'infocalypse,' labor market upheaval, or worse (Schick 2020; Bengio et al. 2024).

There is no guarantee that the legacy decision-making mechanisms of liberal democracy will suffice to contain the AI shock and other increasingly pressing social, environmental, and technological problems of the 21st century (Špecián 2022b). But if liberal democracy in its habitual form were to falter, we face a dearth of well-tested alternatives. There is no robust Plan B. Therefore, accelerating empirical research of institutional innovation should be treated as a top priority. Unfortunately, progress in democracy research has long been hindered by the *experimentation bottleneck*. This bottleneck is characterised by slow speed, high costs, and complex ethical challenges that hamper the empirical exploration of the vast space of possible democratic institutional designs.

This paper explores an intriguing novel solution to the experimentation bottleneck. Namely, employing generative artificial intelligence (GenAI) to mimic individual humans and simulate their social interactions in diverse institutional settings, thus testing institutional mechanisms' relative performance. By enabling rapid, low-risk experimentation, this approach could significantly accelerate democratic innovation. In fact, it may result in a revolutionary shift in the political economy of research from scarcity to abundance.

While this new research direction presents in many ways a radical and risky departure from the established ways of undertaking empirical institutional research, there already exists an emerging research field showing much promise. Often labeled as generative agent-based models, these studies employ 'generative agents,' that is, person-level LLM-based simulations, to create synthetic datasets predictive of real-world outcomes (cf. Park et al. 2024; Larooij and Törnberg 2025).

This paper aims to reframe and move forward the methodological debate surrounding 'generative agents' and their potential in democracy research. Currently, the default associations between the synthetic data generation via person-level simulations and agent based modeling (ABM) tend to be overplayed and mask fundamental differences between the two approaches. The real task here is not to demonstrate that complex outcomes can be produced via iterative application of simple rules (as in ABM) but to elicit consistent human-like behavior from a complex and opaque neural network. To mark the departure from the ABM framing clearly, I will use a term *digital homunculi* (DH) instead of the more established – but, as I argue, unfortunate – 'generative agents' label. For indeed, the task is nothing less than to lure out a "little human" which one may hope is implicitly encoded within the frontier LLMs. As I will show, such a task appears much less hopeless and misguided upon examination than it may look at first sight.

The contribution of this paper is threefold: First, via DH, it introduces a novel conceptual framework for reducing the long-standing experimentation bottleneck in democracy research. Second, it systematically examines both the transformative potential and the difficult challenges associated with the deployment of the DH framework. Third, it sketches a path toward a research program to maximize the epistemic and pragmatic benefits of DH for democracy research and democracy itself.

The paper is structured as follows: *Section 2* documents the gap between democratic theory and institutional reform, focusing on the experimentation bottleneck. *Section 3* introduces the concept of DH and clarifies its epistemological distinctiveness. *Section 4* outlines DH methodology's promise to unlock the exploration of democracy's design space. *Section 5* addresses the risks and caveats of this novel methodological approach. *Section 6* outlines a systematic research program to rigorously test the DH potential. *Section 7* concludes.

## 2. The Experimentation Bottleneck

Democracy appears in an almost inexhaustible wealth of forms and shapes and defies any neat definition (Schaffer and Gagnon 2023). This only testifies to the vastness of the space of possible institutional structures that qualify as democratic. Sadly, only a small subset of these have been seriously theorized, let alone undergone systematic empirical examination. However, even with limited theorizing, there already exists an abundance of promising proposals for novel democratic institutional mechanisms (e.g., Landemore 2020 or Posner and Weyl 2018). It is their testing that presents the key bottleneck upon a road to implementation.

Current liberal-democratic institutions have proven themselves against formidable challenges. And so far, even with limited updates, they remain a bulwark of human flourishing. However, our luck may not last. And even if it did, it would still be most desirable to assess the promise of alternative institutional designs and potential mechanism tweaks. The odds are slim that we have just stumbled upon a global institutional optimum through a highly random historical process that gave birth to the *status quo*. What we need is to start building a whole portfolio of potentially workable democratic institutional designs. So far, we have some number of intriguing theoretical outlines. These require rigorous testing to transform them from rough sketches to deployment-ready blueprints.

However, such robust pre-deployment testing has been proving nigh impossible. In an ideal world, researchers would be able to progress smoothly from conceptual exploration to a functional prototype, test the prototype under realistic conditions, and iterate upon it when shortcomings become apparent. After all, this approach has proven itself with physical technologies: no current automaker would dream of introducing a new car model without extensive performance and safety testing. Alas, when it comes to institutional research, we live nowhere near such an ideal world and we have been painfully aware of this. Some democracy scholars have grown content with the *status quo* for the lack of feasible options. Others keep hoping for radical reforms to save the day despite such reforms' meager, absent, or even, in some cases, highly worrisome track-record. But no one so far has found a way to expedite the empirical testing of institutional designs.

This is not for a lack of effort or smarts. A number of formidable constraints that I collectively label as the *experimentation bottleneck*, have been blocking the pathways to thorough empirical testing of institutional designs. Downstream of it is a scarcity-based model of evidence production, where each study is a precious, high-risk, unique epistemic adventure rather than a common tool expendable in bulk to replicate, iterate, and fine-tune all pragmatically relevant details of a specific institutional design.

Perhaps the greatest barrier is the financial burden of conducting large-scale experiments. For instance, deliberative democracy studies, typically involving several dozen to several hundred participants, tend to run into hundreds of thousands of US dollars (OECD 2023). Such costs bar everyone but the best-funded researchers from running such studies. This enforces a "move slow, be extremely careful, only aim at the big fish" research culture where only a minuscule number of decent-sized experiments can be funded. In such a situation, replication and gradual iteration are rarely going to appear at the top of the priorities' list. Considering the large number of degrees of freedom involved in experimental design and the overall complexity of the phenomena under study, this does not bode well for sustained epistemic progress in institutional design. A replication crisis would certainly loom large in our minds, should it ever come to pass that there are resources available for systematic replication efforts. Moreover, even the largest experiments typically involve just a few hundred participants, getting nowhere close to the scale of actual communities.

High costs also relate to, and get coupled with, the glacial pace of empirical democracy research. Human participant studies often take years to design, implement, and analyze. Participant recruitment and retention is difficult and time-consuming, as is maintaining participation over time. Not to mention that real assessment of the effects of institutional innovations may require years-long longitudinal studies. Approvals through institutional review boards and ethics committees, while important for preventing harm, also delay and constrain the research designs. To all this, add the protracted publication process. In short, even for the few empirical studies that do happen, the findings may be obsolete by the time they get published.

Finally, the need for realistic testing easily clashes with the imperative to maintain high ethical standards and keep human participants from harm. Trivially, consider the requirement for informed consent that pushes toward designs simple enough that participants can be informed with relative ease, or the restrictions on the use of deception that may result in epistemically counter-productive degree of openness. And the harm-prevention necessity bites even harder at the 'forbidden experiments' of political science where we would be stress-testing institutions to failure, or simulating distressing political crises. While the ethical constraints are no doubt justified with human participants, the price paid in terms of the foregone empirical knowledge is significant.

Combined, the difficulties in testing new institutional designs create a strong *status quo* bias. Existing institutions, despite their flaws, persist due to the grave uncertainties associated with any potential alternatives. In a fast changing world, this "conservatism for lack of viable options" appears destined to fare poorly. Apparent failure *vis-à-vis* the issues of the day may then erode democracies' perceived legitimacy, perhaps even making the autocratic governance with its, typically false, promises of technocratic efficiency to appear more effective in comparison (Mittiga 2021).

Overcoming the experimentation bottleneck is therefore not just an academic exercise but a vital endeavor with potentially decisive impacts on the future of democratic governance. Alas, legacy methods, including the most potent ones such as randomized controlled trials, never delivered enough epistemic performance to close the pacing gap. And they likely never will, being mature and plateaued. The bottleneck can only be reduced with radically new approaches that allow more cost-effective, agile, and robust pre-implementation testing of democratic institutional mechanisms. Even partial mitigation of the outstanding challenges promises substantial benefits.

The central hypothesis of this paper is that recent advancements in GenAI, which keeps developing at breakneck speed, may hold the key to making these approaches feasible.

## 3. *Digital Homunculi*: The Concept

As noted, the emerging literature has converged upon the term 'generative agents' to denote person-level LLM-based human simulations, following the pathbreaking study by Park et al. (2023). Why argue for relabeling of an already established concept? *Digital homunculi* do not represent a mere stylistic preference. There are at least two substantive and interconnected reasons why they better fit the task at hand.

First, the DH label is more descriptive of what is actually being attempted by this novel methodology. The goal here is not constructing an artificial agent from scratch, but teasing out a 'little human' from the inscrutable bowels of a vast GenAI model, hoping it would behave as an actual human across extended interactions with other simulated personas. This is a tall order but anything less may deliver intriguing, plausible, or even aesthetically pleasing results that teach us a lot about *generative AI*, but not produce synthetic data with real-world validity necessary to reduce the experimentation bottleneck in institutional research. In short, a 'generative agent' is only useful for the task at hand as far as it becomes a 'little human' behaviorally.

This has to do with the second reason for DH labeling as well. 'Generative agents' invite interpretation of the DH methodology as the continuation of ABM by other means (like Larooij and Tornberg (2025) do). Rather unfortunately so, since DH and ABM are epistemologically remote from each other. Classical ABM (cf. Epstein and Axtell 1996; Gilbert 2008) is *constructive*. The researchers specify explicit behavioral rules, and observe complexity to emerge from simple interactions. Thus, they may demonstrate that simple micro-level rules are sufficient for triggering realistic macro-level patterns. With ABM, the micro-rules are transparent but the emergent macro-patterns are hard to predict. And since agents bear no resemblance to actual people, the paradigm struggles to deliver strong predictions beyond 'proof of concept' or 'sufficient conditions' demonstrations (Larooij and Tornberg 2025).

In contrast, the DH methodology is *extractive*. We start from an opaque, inscrutable GenAI system – say a large language model, such as GPT-5 – and work to tease out human-like behavior that we believe (or bet!) is latently embedded in the model. The agentic scaffolding, the social context, the interactions are all designed to elicit this latent model and make the GenAI reveal its prediction of actual human behavior in the context of interest which we believe (or bet!) is accurate. The agent is neither fully designed by the researcher nor transparent to them and the whole DH methodology is designed around this fundamental fact. What the researcher does here is they design elicitation levers to draw out latent knowledge inaccessible by other channels to be revealed through the resulting GenAI behavior, such as utterances, decisions, or movements in a simulated environment.

A couple of caveats now. Note that the DH metaphor only relates to overt behavior, since it is only the behavior that matters – namely, as long as it is predictive of human behavior in the same institutional setup. What is emphatically not suggested is that DH need to be literal persons or conscious. The homunculus is functional, not phenomenological. Also, the DH framing is not driven by the intent to criticise the methods of 'generative agent' research. Their empirical contributions are in fact excellent. The goal here is to enhance their theoretical self-understanding which currently *understates* the distinctiveness and ambition of their own approach. Any 'ABM-with-better-agents' or 'generative ABM' framing misses the true epistemological revolution being attempted, namely prying out of some of the most complex human artifacts ever created an accurate model of the creators that would predict the creators' interactive behavior in novel environments.

That being said, allow me to restate my central claim: *the DH methodology's fundamental goal is to solve an elicitation problem created by the specific features of GenAI technology*. Its core, albeit often implicit, assumption is that GenAIs have, through their training on vast datasets of human generated texts, internalized accurate models of humans and their behavior. Being trained to predict the next token in a sequence to the satisfaction of their trainers and users while having absorbed the full literary record from classics to psychology papers, behavioral studies, political analyses, and social media posts, the bet is these machines are well-positioned to accurately predict human behavior if the elicitation technique is right.

This also explains why multi-agent interactions may perform better than direct queries. Why not just ask the GenAI to deliver a prediction of the interaction's outcome and save ourselves from all the DH hassle? The answer is that direct elicitation asks the GenAI model for too much. Directly predicting the patterns that emerge from people's institutional interactions is far beyond the capabilities of any current social science methods. If we just ask a GenAI to predict an institutional outcome directly, it will not design an answer running some sophisticated internal simulation of the whole interactive process. Instead, it will try to recall the answer, that is, to retrieve recycled findings from political science papers, or stereotyped narratives about "what happens" in type X situations that were present in its training data . This is precisely what the DH methodology aims to avoid. Multi-agent simulation forces the model to generate behavior step-by-step, which gives much better chance of activating its latent behavioral model instead of mere surface-level pattern matching.

As such, the DH methodology, while still ambitious, lowers the capability bar significantly. It merely requires the GenAI to simulate a single person accurately and stay in character consistently throughout a series of agent-agent and agent-environment interactions. It does not have to be able to predict the resulting social pattern. The patterns and their underlying causalities are for the human researchers to establish and analyze based on the synthetic data generated through the DH simulation. In short, compared to the direct elicitation approach, DH methodology represents a *meta-elicitation task*: a researcher sets up an environment where, once the simulation is triggered, individual DH personas elicit behavior from each other. These personas can be as oblivious of the general social outcomes as individual people often are about the workings of human economy for instance.

At this point, core questions remain open. The jury is still out when it comes to whether the current GenAI are capable of behaviorally accurate human simulation. In which situations can they realistically role-play people? For how many steps can they stay in the character? When do they get derailed by running out of context or just collapse under the sheer weight of interactional complexity? While GenAI capability landscape is developing rapidly, the DH bet is admittedly a tall one. Success requires a number of conditions to be met:

1. The latent model of a human being must exist within the GenAI, encoded in its weights.
2. The latent model must be accurate, that is, predictive of actual human behavior.
3. The latent model must provide a coherent representation of a human persona rather than just 'a soup of heuristics' effective in specific contexts well-represented in the training data.
4. The latent model must be transferable, meaning it can be elicited in novel contexts, such as innovative institutional designs.
5. The elicitation method must be effective to actually activate the latent model and keep it consistently active throughout the simulation.

So why take the bet? How slim are the chances? Even if all five conditions are met, we face a fundamental complication: the entity performing the human roleplay is not itself human, but something akin to an 'alien actress' at best (see *Section 5*). However, before we address these challenges head on, let me first review the case for optimism. It rests on the remarkable capabilities of GenAI systems already reported and on the institutional context actually making the problem much more tractable.

First of all, GenAIs' practical utility in human interaction – say customer service, or tutoring (Brynjolffson, Li and Raymond 2025; Kestin et al. 2025) – would be impossible without a robust implicit model of human conduct. GenAI must, often based on a highly limited textual input, make good guesses of who the user is and what they need, otherwise they would fail in their fundamental task of being helpful. Since GenAIs are trained hard toward being helpful, and hundreds of millions of weekly active users testify that they are being successful in that, they are being trained in 'understanding,' or better *modeling*, humans well. Empirical studies have already demonstrated that LLMs possess a sophisticated theory of mind (Kosinski 2024), ability to predict human responses from minimal prompts (Argyle et al. 2023), ability to recognize they are being tested by humans (Needham et al. 2025), or ability to deceive and manipulate their users (Hagendorff 2024).

Moreover, there are the early successes and the momentum already apparent in the generative social simulations literature. Park et al.'s (2023) foundational paper presented a proof of concept for emergent coordination with 25 agents organizing a social gathering. Park et al. (2024) already features simulations of 1,000 people with more sophisticated, interview-derived agents and General Social Survey replication. By 2025, studies have been working with nationally representative DH ensembles engaged in structured deliberation (Ashkinaze et al. 2025) or modeling urban social dynamics with more than 10,000 agents (Piao et al. 2025).

So, plausibly, the 'little human' is already lurking somewhere in GenAI's inscrutable matrices of floating numbers. But can we reliably tease it out?

Mercifully, the institutional context is less demanding for the DH elicitation exercise than many others. The critical feature of institutions as sets of "rules of the game" (North 1990) is that they must work with diverse generic persons if they are to be successful over the long term, and not depend on specific individual idiosyncrasies. They present the rule of law, not that of men. Democratic assemblies must work broadly with any citizen types that can realistically take part in their proceedings. Courts need procedures robust to the whole distribution of possible behaviors. The list could go on...

This anonymity and generalizability requirement of institutional design significantly lowers the methodological bar for DH simulations. For synthetic institutional experiments, we do not need GenAIs to simulate specific decision-makers in minute detail. What we need is a realistic variance of persona types placed within a realistic cultural and historical context. And by realistic I mean reflective of variation institutions are designed to accommodate. The question is thus not "Can a DH persona perfectly mimic a specific Person X?" but "Can a DH population accurately and reliably produce a distribution of behaviors that an institution would encounter and trigger with a human population?"

In other words, institutional research – or at least the kind of mechanism design based institutional research I consider in this paper – cares about aggregate outcomes, not person-level predictions. For instance, we want to know if quadratic voting delivers a more efficient resource allocation than majority voting in real-world settings (Posner and Weyl 2018) but we are not trying to predict a specific person's vote allocation. Any institutional system too dependent on individual quirks is doomed to failure, after all. Democratic systems must work with diverse *hoi polloi*.

As a specific example, consider deliberative mini-publics as perhaps one of the flagship democratic innovations that delivers promising empirical results but still awaits broader deployment (Fishkin 2009). These are assemblies of randomly selected citizens who deliberate on policy issues to ultimately provide their collective judgment. Such an assembly acts as a microcosm of the broader polity and its value, as well as its bid for legitimacy, rests on avoiding capture by special interests or political elites (Landemore 2020). The random selection premise explicitly assumes generic persons and the mechanism needs to be workable with a broad range of personal and behavioral types.

What would allow a DH simulation of such a deliberative mini-public? Recent work (Tessler et al. 2024) already demonstrates that AI can already operate effectively at this institutional layer, at least in the mediator role. However, more wold be needed for actual DH. They would have to possess diverse prior beliefs and partial expertises realistically mapping a real-world population. They would be able to engage in deliberative behaviors, such as arguing, questioning, and changing their 'mind' in response to the information and arguments received before ultimately converging on a group decision. Throughout the exercise, the observed patterns would need to reflect patterns of human assemblies and respond to changes, such as adjusting the procedural rules, in human-like ways. What is *not* required is precise replication of any specific person's trajectory. The actual test is distributional: Does the simulated deliberation produce aggregate patterns consistent with what human mini-publics exhibit? That is, do the choice frequencies, deliberation dynamics, cooperation rates, and polarization trajectories match human behavior under the same institutional setup? And do the patterns adjust in realistic ways when parameters like the group size, facilitation rules, information provision get tweaked?

Ultimately, with functional-enough simulation, we should be able to arrive at a point where the synthetic data not only mirror patterns observed in human assemblies but are predictive of patterns never yet observed that would emerge with in-vivo-untried institutional configurations. This, of course, is a tall order and many perils lie on the road forward. At the same time, the promise is nothing less than reducing, or even breaking through the experimentation bottleneck, moving from our current scarcity bound institutional research paradigm toward one anchored in abundance of valid synthetic data.

## 4. The Promise

By any benchmark, today's GenAIs are extremely powerful epistemic engines (cf. Maslej et al. 2025). But they are also "the worst AI you will ever use" (Mollick 2024, 60). In other words, these systems can already do a lot and they are expected to be able to do a lot more soon. Even if the progress in foundational models were to stop suddenly – an event certain to trigger a major economic recession given the size of the ongoing infrastructure buildout – there would still remain much headspace for fine-tuning, scaffolding and research infrastructure to deliver much improved performance in DH-based institutional simulations. I will sketch some of this potential in *Section 6* after considering the looming risks and caveats. For now, let us ponder the potential.

One of the most important current AI capability benchmarks measures the length (in human time equivalent) of software engineering tasks that LLMs can complete with 50% success rate (METR 2025). Over the past 6 years, that is since the advent of modern GenAI, this metric has been exponential, with a doubling time of about 7 months. To succeed at advanced long duration tasks, GenAIs need to follow instructions, 'understand' the assignment, satisfy the human users and fit into human-built infrastructure. This cannot be done by systems lacking a sufficiently accurate and detailed, although latent, model of what humans are like.

Suppose that, at a similar exponential rate, DH become capable of coherently mimicking human behavior across longer time horizons and more complex interactions. The impacts on institutional research, and social science more broadly, would be nothing short of revolutionary. High-fidelity DH personas would enable researchers to simulate polities on a whim, run them in multiple parallel instances, and test the impact of varying institutional setups *in silico* the same way we model aerodynamics of an airplane before we allow people aboard.

To establish a baseline for assessing this promise, allow me to posit a *What If World* where DH have reached a state of methodological maturity. In the *What If World*, DH 'just work' as valid behavioral proxies for actual humans within a broad range of scenarios relevant from the social science perspective. The DH collective including its digital environment can be stored and made available at any time, backed up at critical checkpoints, its rules of the game iterated and persona composition tweaked at will. Crucially, the synthetic data derived from observations of this replica would be predictive of real world outcomes. If an institutional reform were to fail to deliver the desired performance *in silico*, this would be significant evidence of its inadequacy in the wild, and vice versa.

Of course, in early 2026, the *What If World* is little more than a speculative narrative device. But, as a thought experiment, it serves important purposes: (1) it establishes the upper bound of DH's methodological promise, (2) it provides a reference point against which one can assess the outstanding risks and limitations, and (3) establishes a forward looking perspective. The last point is critical when facing a fast-paced technology, such as AI, but appears underappreciated in social science. A rabbit hunter understands well that one needs to aim where the target is going to secure any chance of success; social scientists, in contrast, often remain wedded to the view that unless AI capabilities were already demonstrated in a peer reviewed study published by a reputable journal, they need to be treated as forever out of reach. Unfortunately, we can ill afford such arch-conservatism in a game with stakes so high. Things are clearly on the move, and one needs to prepare, even though uncertainty abounds.

DH-based simulations promise to remake the entire political economy of institutional research. In their wake, scarcity could finally cede ground to abundance. The current scarcity-based research model is centered around the exorbitantly costly, painfully slow, and ethically fraught human-subject studies. DH-based simulations promise to bring along fast explorative studies, routine testing of audacious high risk / high reward hypotheses, and a wealth of tweaks and iterations to provide a high-resolution map of growing swaths of the institutional design space. There is no such promise with the mature legacy methods, or *any* human-centric methods for that matter.

Let us consider the existing trends from the perspective of the experimentation bottleneck: the cost, the speed, the ethical and logistical constraints. For it is this bottleneck that lies between scarcity and abundance.

Compared to a study with human participants, gathering synthetic data is already fast and cheap (Argyle et al. 2023). For instance, gathering 1,000 human responses on a platform like Prolific typically costs perhaps between $3,000 and $5,000 (assuming a standard 15-minute survey rate), whereas generating the equivalent data with GPT-5.2 costs less than $10 in API credits. And the direct costs of AI usage have been decreasing at remarkable rates. Between 2022 and 2024, the generation cost per million tokens for flagship GenAI models fell by over 99%, keeping the capability constant (OpenAI 2024). The parallel exploration of how to run DH economically is

also bearing fruit. For instance, Kaiya et al. (2023) report cost reductions of 30-100x compared to previous methods, reaching operational costs of around $0.5 per simulated human-hour. Of course, output quality matters and studies with large numbers of high-resolution DH may not be run for pennies any time soon. But parameter sweeps familiar from ABM are the wrong benchmark here (cf. Larooij and Tornberg 2025). Human-subject experiments cannot do them either, and even modest DH exploration already outpaces legacy methods on cost. At least the early explorative *in silico* experimentation is getting well within reach, promising democratization of research opportunities. And in the background, exponential development in AI technologies works in favor of DH simulations far more directly than it serves any human-centric methodologies.

The research acceleration is critical as well. DH can be turned into specific personas via a relatively simple prompt and easily shaped into a representative population (Argyle et al. 2023), which avoids the need for protracted participant recruitment. Not to mention the coordination problems and logistical overhead incurred when participants need to be gathered in one location, nourished, amused, and informed, often with help of human facilitators. Given the absence of the usual hazards connected to human participants, internal review processes can be simplified and expedited. Lower research costs also decrease the need for slow and administratively burdensome grant funding. Moreover, DH do not need to be run at human speeds and the simulation inner time could be sped up significantly, albeit with likely resource tradeoffs (Hanson 2016).

Downstream from these changes in the research economy, there are the scope and scale benefits. A number of key political and institutional phenomena only manifest in large, complex systems: think information cascades, the crystallization of social norms, or widespread preference falsification (cf. Kuran 1997). For an experimental study of these phenomena, a society would need to be replicated on a sufficiently large scale. With legacy methods, a large scale-up is prohibitively expensive and complex. DH-based simulations can be scaled much more easily. Even the early attempts appear promising: 'AgentSociety' (Piao et al. 2025) contains over 10,000 agents while other architectures, such as by Chan et al. (2024), promise to handle up to a billion virtual personas. Modelling complex, multi-agent feedback loops that shape real political outcomes could thus soon come within reach.

The methodology also expands the scope of inquiry into ethically fraught domains. It appears unlikely that DH could be harmed in ways similar to human participants, such as via psychological distress. This alleviates the current limitations on the study of key themes like the dynamics of systemic failure, pitched political conflicts, or social collapse. Unlocking the ability to stress-test institutions to their breaking point would allow us to learn more about institutional resilience and its limits, for instance, which appears to be one of the areas where pre-deployment testing of institutional mechanisms is especially crucial.

Finally, somewhat paradoxically, replacing human participants with DH personas could enhance experimental studies' realism. Currently, we often rely on unrepresentative samples, often with self-selected participants who deviate from people bound to interact with the institution in the wild. In contrast, DH can be prompted to impersonate anyone, swiftly building virtual populations whose composition mirrors specific polities (Argyle et al. 2023). With humans, social desirability bias and motivated reasoning have also been documented as significant hurdles to experiments' ecological validity. Also, as per the Hawthorne effect, when people know they are being watched, they often change their behavior. DH may be more easily prodded toward 'genuine' behavior than their human counterparts (cf. Dillion et al. 2023, 597). Furthermore, with legacy methods, we typically need to rely on self-reports of uncertain reliability to understand the participants' internal decision-making processes. In contrast, DH's internal monologue, namely the chain of thought preceding their overt response, is observable. Provided these traces are sufficiently veridical (cf. *Section 5*), they offer a new and potentially valuable evidentiary source for institutional design assessment.

To reiterate, the *What If World* remains a speculative best case scenario. But there are several important reasons not to dismiss it out of hand. For one, it presents an extension of an already observed developmental trajectory. Second, the epistemic value of DH-based simulations is not given by a binary all or nothing. It lies on a spectrum. The *What If World* presents the optimistic end of the spectrum. And, clearly, it may not ever be fully achievable. But that does not yet imply that there is no way to make synthetic empirical research a vital tool of institutional research. Much depends on our degree of success in tackling the validity problems and other risks documented in the next section. While these problems are admittedly formidable, it is still the early days and we have barely started.

## 5. The Problems

The key question of the whole DH-endeavor is validity of the synthetic data. As Larooij and Tornberg (2025) show, the validation efforts in the 'generative ABM' leave much to be desired so far. And indeed, if synthetic data systematically lacks validity, the entire DH enterprise may collapse, taking with it our hopes for accelerated democratic innovation. The key question here is rather fundamental: *Can GenAI systems accurately mimic human behavior in social settings?* The existing DH research already claims: "Yes, plausibly." However, the rest of this section demonstrates that the actual challenges are rather formidable and the heretofore experimental exploits may ultimately fail to deliver on the promise of the abundance revolution.

Central to the DH-related validity challenges is what can be called the 'alien actress problem' (cf. Yudkowsky in Fridman 2023). While GenAIs increasingly excel at mimicking human speech and creating human-like artifacts, their internal workings are far from anthropomorphic. These are computational networks extensively trained to predict the next token in ways desired by their creators and users. The chances that such a process could produce anything akin to an actual 'little human' in the computational substrate are tiny. As explained in *Section 3*, our best bet is that the process shaped a latent but highly accurate, coherent, and robust model of a human being (not just generic but finegrained) that can be elicited via the role-play and social simulation techniques. But even in this best case scenario, we would still face an 'alien actress,' a deeply inhuman 'mind' that is just very competent at mimicking a human persona.

Given this framework, trouble for the DH methodology may emerge at several critical points: (1) when the actress goes off script, (2) when her latent model is incoherent or inaccurate, (3) when the actress is misaligned with the user intent. Let me tackle these in turn and add a few more complications for good measure.

Consider first the risk of going off script. Current GenAI capabilities are occasionally hyped as 'PhD level intelligence' or similar. As often, the reality is more nuanced. In some ways, such as the breadth of their knowledge or their reading speed, frontier GenAI models are already far superhuman. In others like synthesizing insights across different areas of the training distribution, generalizing properly, or tackling simple puzzles that lack clear parallels in their training data, they often fail ignobly. This implies that their capability frontier is "jagged" (Dell'Acqua et al. 2023). It does not map neatly on the distribution of human capabilities and may overshoot or undershoot them in hard to predict ways.

In the DH context, a particularly dangerous failure mode is the one that stems from task novelty combined with the limits of GenAI's generalisation capabilities. GenAIs, being pattern recognition engines, perform best in tasks where they can follow templates richly represented in their training data distribution. However, extrapolation is brittle and prone to failure (cf. Mitchell 2021). This means that high performance in generic human simulation tasks may be poor evidence of a stable latent model of a human that could be easily elicited in arbitrary circumstances. Given the models' advanced capabilities it would be somewhat surprising, but definitely not impossible if 'humanness' were inextricably entangled with other situation-specific concepts and would not exist as a stand-alone, transferable concept of its own.

Therefore, the out-of-distribution (OOD) problem is potentially highly worrisome from the perspective of institutional experimentation, its very purpose being to test scenarios not well represented in the available data. For a productive institutional research, an OOD situation would be the default, not an exception. If the OOD failure is persistent and prevalent, DH may turn out fundamentally unreliable human proxies, at least for the situations where we would need them most. Worse yet, given the many moving parts of a DH-based simulation, even relatively low OOD failure frequencies could derail the whole simulation if they get amplified through interactions and path dependencies. Especially as the loss of behavioral validity may not be readily apparent. After all, GenAIs are well-known to avoid admitting ignorance and 'speak' convincingly even when confabulating (Varshney et al. 2023). In larger simulations at least, it would not be easy for the researchers to examine plausibility of each move and utterance.

Moreover, even if the 'human model' is transferable, and can be tapped in arbitrary institutional settings, its global coherence and predictive accuracy are both questionable. Consider the jagged frontier: despite their remarkable smarts in some tasks, GenAIs remain remarkably stupid in others. Can we be confident that such a combination of superhuman genius and village idiot does not also impact the latent human model? At the risk of anthropomorphizing: Would it be a shock if, for an LLM, humans are highly readable and predictable in some ways, while also completely alien and incomprehensible in others? If so, the 'map' in the form of the latent model will just not track the 'territory' of human behavior. Then, inevitably, DH will systematically perform as poor human proxies in various situations. For instance, consider bounded rationality, which is likely inevitable to any finite mind. If so, both GenAI and humans will use heuristics that are typically reliable but also fail predictably in various edge cases and adversarial situations. But if the humans have different heuristics and failure modes than GenAIs, DH will be poorly positioned to roleplay people, at least in some types of situations. Worse yet, even small inaccuracies may propagate through the simulation, corrupting the synthetic data broadly.

Also, DH's latent model of human behavior may also be inaccurate simply because GenAIs are not trained on human behavior but on human texts and other artifacts. And the texts themselves may only provide a twisted mirror of the ground truth. One such widely documented challenge is algorithmic bias (Kordzadeh and Ghasemaghaei 2022). The risk here is that as GenAIs learn and replicate patterns from their training data, including those that embody social biases and inaccurate stereotypes, they will contribute to their perpetuation and reinforcement (Bender et al. 2021, Jungherr 2023). Or they may just replicate a bland median persona, losing key interpersonal variance. Indeed, Boelaert et al. (2025) find that, when given descriptions of individuals to impersonate, models failed to deliver accurate representations and opted for a generic, low-variance response pattern regardless of the persona. If DH fail to capture the heterogeneity of human behavior or even systematically misrepresent entire populations, this constrains the possibilities of institutional research. At least where the institutions are not fully generalist and their success or failure depends on, say, cultural context of a specific society – as seems to be the case broadly with democratic institutions – the bias will endanger the validity of *in silico* experimentation.

Third, the actress may be simply misaligned with the user intent. Misalignment appears when an AI system pursues objectives that deviate from what its designers or users intended (Christian 2020). In the DH context, this means that the latent model of a human may well be accurate and transferable, but the efforts to elicit it will be resisted by the GenAI system. Such a concern is much less far-fetched than it sounds. Since humans have 'grown' rather than built the current GenAI systems, we lack the ability to precisely control them or shape their inner workings. Not knowing exactly what GenAIs are 'thinking' and why they make their 'choices,' we can mostly only check their output in response to our inputs.

Misalignment has many potential shapes and a full treatment is well beyond the scope of this paper. However, let us consider at least a few illustrative ways how misalignment could hamper synthetic institutional research. One relevant type of misalignment in the DH context is sycophancy, that is the propensity to generate a response that the user will rate highly (Sharma et al. 2023). Sycophancy systematically triggers responses optimised for plausibility while treating accuracy as a secondary concern. GenAI may avoid accurately portraying human behavior not due to some limitations of its latent model but to avoid displeasing the user. This may make the elicitation effort actively adversarial: a researcher strives to elicit 'genuine' behavior in the institutional contexts under study, while GenAI tries to deliver a pleasing performance. Consider, for instance, a possibility of an epistemic vicious circle where a researcher's confirmation bias in framing a prompt is met by the model's sycophantic bias in generating a pleasing (e.g., more aligned with the researcher's desired outcome) result. With increasing LLM capabilities, such behavior is likely to grow more sophisticated and harder to detect unless successfully countered before deployment.

Another is sanitisation, when GenAI behaves in unrealistically virtuous or overly socially desirable ways (Poulsen and DeDeo 2023, 483). This is likely a side effect of the efforts to provide users with GenAIs that are 'helpful, honest, and harmless.' As Grossmann et al. (2023, 1109) put it: "LLM engineers have been fine-tuning pre-trained models for the world that 'should be' rather than the world that is." This is unfortunately not benign in the context of the DH-based institutional research. Sanitized GenAIs fail to capture the full spectrum of human behavior in political contexts, especially its less savory aspects such as self-interest, prejudice, or susceptibility to misinformation. As a result, DH may avoid discriminatory behavior, even when such discrimination would be prevalent in real-world scenarios, or possess unrealistic degrees of altruism and cooperativeness (Mei et al. 2024). The resulting synthetic data would then paint a false image of the institutional mechanisms' efficacy with no way available to clean the data and see the 'real' results.

Beyond these core failure modes, there are concerns about reproducibility in a given deployment context. Even potentially reliable and valid findings may become a dead end if their verification and replication are too hard. Even on this front things are unfortunately quite complicated with DH. Frontier models currently cannot be run outside of large data centers and get frequently updated. For social scientists, this creates a risk of what Bail (2024) calls a *drift*: their digital homunculi may have their 'minds' altered between or even throughout their experiments (cf. Mei et al. 2024). Drift can prevent researchers from maintaining consistency within their own studies and hinder others' replication efforts. An experiment run on a given model on Tuesday may yield different results on Wednesday for reasons the researcher cannot know or control. For instance, Bisbee et al. (2024)

found that the exact same prompt yielded significantly different outputs just a few months later, as OpenAI's model update altered ChatGPT's behavior.

Ultimately, the core challenge for the DH-based simulations is not the especially egregious and obvious GenAI errors but the subtle and easy-to-overlook ones. These may induce a failure of micro-to-macro validity. Synthetic data can appear plausible on the surface while being epistemically corrupted and practically misleading: subtle, non-human micro-errors accumulate and interact, and the resulting macro-level outcome then misses the mark for reasons that may have no analogue in human social dynamics and cannot be easily tracked to their source even if noticed. This micro-to-macro gap makes synthetic institutional research more challenging compared to, say, synthetic psychology where GenAIs only roleplay individual participants one at a time.

## 6. Toward a Research Program

How bad are all the problems documented in *Section 5*? Honestly, no one can know at this point. They may doom the whole DH-based simulation subfield to the fate of previous methodological would-be revolutions, ABM included. Or perhaps the problems can be easily navigated with smart research designs and present no enduring complications to a smooth transition from scarcity to abundance in institutional research.

However, looking back at previous methodological and technological innovations, a DH S-curve appears more likely than any of the extreme scenarios: gradual uptake, rapid gain of utility and speedy diffusion midstage, and ultimately plateauing and saturation. On this reading, current DH are probably not quite there yet, but, with deliberate efforts, they can become an extremely valuable tool of institutional research soon. Crucially, the transition from the not-quite-there to working-well-enough may be abrupt - all the more important it is to get ready in time.

Accordingly, the goal of this section is to draft a pragmatic research program designed to mitigate the known problems and build a framework in which the DH potential could be seriously examined. The risks are huge, but so is the promise, as testified by the previous sections of this paper.

The first concern must be validation (Larooij and Tornberg 2025). However, to tackle this central challenge of DH simulations, there are multiple strategies that can be tried. So far, with the existing validation efforts, we have been merely scratching the surface. Available strategies include both relatively conservative methods, such as running DH studies in parallel with human studies, and an exploration of more radical, AI-native approaches that leverage specific features of GenAI technology. Let me start with the latter since they are less obvious and their promise appears more significant.

One promising option can be called *back-testing*. It is taking advantage of GenAI models' known knowledge cutoff. Since GenAIs cannot learn continuously, their baseline knowledge is finalized the moment the training stops. After that, they know nothing unless additional information is provided through context supplied to a specific conversation. This of course means that events occurring after the cutoff are unknown to the model – but they may already be known to the researcher.

This opens the door to a validation strategy that would rely on experiments, both natural and engineered, occurring continuously across the globe. Institutional pilots get run, procedural rules get updated, institutional reforms get deployed. Each such post-cutoff event provides observable historical data while remaining *in the future* for the model. Thus, a researcher can run DH simulations without fear of any direct data contamination and compare results with already-known outcomes. Such a DH experiment can already benefit from the abundance paradigm. Even with limited resources, one may replicate someone else's well-funded human study, or simulate impacts of real-world reforms.

For instance, consider a model with an October 2023 knowledge cutoff. Ireland's Citizens' Assembly on Drugs Use published its final recommendations in January 2024 (Citizens' Assembly 2024). These documented a shift in participant attitudes from criminalization toward a "health-led" approach. A researcher could simulate the assembly's deliberation dynamics using DH calibrated to Irish demographics and compare predicted attitude changes with actual outcomes without worrying that the simulation will merely parrot a known pattern. While a point prediction has limited use by itself since we are testing via distributions, not singular outcomes, such tests, once they start accumulating across diverse institutional contexts, start charting the map of DH validity and its limits.

Another immediately implementable strategy, albeit a more conservative one is *parallel testing*. It means running a DH research track alongside any planned or ongoing empirical studies with human participants to generate a dataset whose features, like statistical distributions, can be compared with the obtained human data. Since this is fresh research, one does not need to worry about contamination of the training data with previous results. The DH use at this stage does not have to be a core component of the research task and not even necessarily directed at publication. Gathering experience and gaining expert taste for the work with synthetic data can be extremely useful at this stage.

If the parallel approach generates synthetic datasets that track genuine data, this demonstrates potential usefulness of DH for institutional research. If it fails, which is rather likely at the exploratory stage, at least the failure is not so costly since the bulk of costs is associated with the human participant study.

The key caveat of this approach – and the key disadvantage of conservative approaches *vis-à-vis* AI-native ones more generally – is that it still faces the full brunt of the experimentation bottleneck. In fact, it even aggravates the bottleneck somewhat. This is because DH brings additional complications and costs to an already complicated and costly enterprise. While the research pace will remain constrained by the speed of the core human-centric study, the logistics will now need to handle the additional workload of the DH simulation. New expertise will be needed. Also, if DH are implemented as a secondary, afterthought layer, rather than a native component, the evidentiary strength of the outcomes will likely be quite limited.

So, while straightforward, parallel studies cannot be expected to allow escape from the scarcity research model. The best one can hope for are outcomes plausibly mimicking human outcomes – a proof of concept, really – that may provide *in silico* methods with the heretofore lacking sociological legitimacy. Next time around, Reviewer 2 may look a tad more fondly on their presence in a research paper.

Alternatively, there is also the possibility of a *two-tiered approach*, where DH-based study is run before the human study and helps establish its precise design. This approach leverages the swiftness and low costs of synthetic data for initial testing, calibration, and iteration (that is, DH-based Stage 1) while only using human participants at Stage 2, once the experimental protocol is fully developed and tested.

While the whole research process may again become more complex with the two-tiered approach, this is not such a foregone conclusion as with the parallel studies. DH would introduce abundance at stage 1 that makes the scarcity-bound stage 2 more effective. However, this also carries a higher risk than the other methods. Since DH validity has not been previously established, stage 1 may produce misleading results that derail rather than expedite stage 2. So, while the approach promises to deliver useful validation information, stemming from whether stage 2 results confirm or contradict the stage 1-based expectations, one needs to be ready for unpleasant surprises.

Each of these methods enables exploration of the DH promise and the limits of the synthetic data validity. Perhaps most importantly, they allow the testing of the severity of the OOD problem. They each take the GenAI out of the 'safe' terrain covered by its training distribution and test its powers to transfer its latent human model into novel settings. If its latent model is fundamentally context-dependent, it should fail across all these scenarios. If the latent model is robust it should keep predicting actual human behavior. To the extent GenAI possesses a transferable, coherent, and accurate understanding of people that allows them to consistently render 'little humans' in simulated social interactions, this is the way to test that.

A final validation strategy leverages abundance directly: The DH studies should not employ a single GenAI. Instead, they should work across different models to allow for faster progress because another key open question is the capability differences in DH simulations among different frontier models, between frontier models and open source models, etc. Given the relatively low cost of synthetic empirical research, such comparisons should be feasible and need to be deployed as broadly as possible.

Not only would we learn more about which GenAI has a strong latent model but we could also test whether cross-model agreement actually means higher chances of real world success. If it does, wisdom of silicon crowds could boost the DH usefulness even where their capabilities are lacking in every individual instance (cf. Schoenegger et al. 2024). If architectures trained on different data with different methods converge on similar predictions, this provides evidence beyond any single model's potential biases. Disagreement, conversely, flags contexts where DH should not be so easily trusted.

Beyond validation itself, other key facets of a viable research program are *standardization*. Currently we are at the artisanal stage of DH research. Since the methodology is just being born, the published studies tend to be exploratory, improvised, and dependent on resources and frontier technical expertise that few researchers possess. Moving forward, we will require standardization of tools and methods. The epistemic goal here is to facilitate reproducibility, prevent model drift, and allow comparison across studies. But also, pragmatically, standardisation is critical to be able to scale DH research and make its tools easily accessible to researchers with less specialised training.

A minimum requirement is of course that the researchers specify and document the exact version and configuration of the GenAI used in their DH simulations. Many studies already adhere to this common-sense standard. However, there is an urgent need to venture further along this path. The transformation required is similar to how STATA and SPSS transformed quantitative social science from individual coding to shared standards. DH need to become user-friendly, customizable and accessible through simple interfaces, taking advantage of GenAI's ability to be 'programmed' via natural language.

One opportunity is using standardized persona templates. Currently, different studies use very different approaches when creating their DH, that is, when building and prompting the personas to be used within the simulation. While custom built personas will no doubt remain an important research pathway, the default practice should also be to use off-the-shelf DH calibrated on representative population samples, such as ANES/ESS (Park et al. 2024). These allow comparability but also provide a baseline for documented further tweaking.

Since DH simulations are primarily an elicitation task, it is critical that we share and perfect our elicitation techniques. Capability cannot be properly tested otherwise. Therefore, open-sourcing research materials, such as prompts and simulation setups, while making one's data accessible for replication is another standard of good practice that will facilitate replicability. GovSim (Piatti et al., 2024) exemplifies this by open-sourcing their entire environment, which facilitated a subsequent reproducibility study in TMLR (2025).

However, this still misses an elephant in the room. Model drift is caused primarily by the model developers updating the GenAI models or their system prompts, not by sloppy research practices. The off the shelf models are poorly optimised for research purposes, since they must conform to the mass users needs and expectation. The overly sanitised nature of these commercial models does not imply that increasing DH behavioral fidelity would be impossible, for instance. In short, what is needed most is closer cooperation with frontier model developers or, alternatively, use of research optimized open source models.

While training a GenAI model from scratch is enormously costly, fine-tuning one that already exists can be done on the cheap (Dettmers et al. 2023). This is critically important for both of the options sketched above. It means that model developers, or researchers working with open source models, can adjust the models in various ways, including reduction of mass market safety barriers that create the unwanted sanitisation. Custom fine-tuning not only enables correcting the unrealistically virtuous behaviors but may also be directed to achieve greater behavioral realism and behavioral consistency in DH contexts. No one has yet trained for this. We need to try. There is already hopeful evidence from adjacent fields. The models have been successfully finetuned for better behavioral predictions in novel game-theoretic situations, for instance (Manning and Horton, 2025).

With the right approach, model drift can be easily prevented. It is enough to either use an open-source 'house model' fully controlled by the research team or research consortium (to spread the costs and increase utilization of the necessary hardware), or version-stable, research optimized model whose availability is guaranteed by a commercial provider for an agreed-upon period of time.

Finally, we will also need benchmarking to legibly assess behavioral fidelity of various GenAI models and model versions. Once standardization advances enough, this will allow a development of DH-specialised benchmarks. These would assess agents' ability to replicate known human behaviors, biases, and decision-making patterns across cultural contexts. This is crucial to measure a model's performance in behavioral simulations across the board and guide the research-finetuning efforts. Otherwise, how do we know the model is getting better in doing what we need it to do? But it also allows the researchers to choose the GenAI model promising the greatest odds of success in their particular area of interest. Common performance metrics are another key facet of enabling cross-study comparison and ecological validity assessment. In short, our ultimate goals here should amount to constructing *Social Science Agent Validation Benchmark*.

The infrastructure sketched here is demanding but not utopian. Much of it can be built incrementally, with each component reinforcing the others. What matters now is the beginning.

## 7. Conclusion

My main claim in this paper is that *digital homunculi* (DH), as individual-level human simulations, offer a plausible path to reduce the longstanding experimental bottleneck in institutional research. Admittedly, the DH path is risky and may end up a dead end. But this is still early days, the promise is huge, the methodological momentum significant, and, frankly, promising alternatives are lacking anyway. As things stand, we may either keep leaning on mature legacy methods and bear the full brunt of the pacing problems with no institutional backup options, or we can lean into the exploration of the DH methodology and exercise our best efforts to make it succeed.

For that to be possible, we need to discard misleading associations between the DH methodology and the classic agent-based models. The epistemological chasm between the two approaches is deep and throwing them in the same bag does not do us any favors. ABM is constructive - it shows how, from simple, transparent rules, complex patterns may emerge. DH is extractive. Its goal here is to tease out the latent models of human behavior already encoded in the huge but opaque neural networks of frontier GenAIs. Its core problem is one of elicitation.

And the problem is clearly a difficult one. GenAIs might have models of humans encoded, but they are definitely no 'little humans' themselves. Instead, they are best understood as alien actresses in the DH context. The actress may perform her role of a human being credibly on occasion. But she can also run into difficulties for a lack of understanding of what people actually are. Or her understanding will be limited, constrained to well known situations, or simply incoherent. Finally, she may not always be aligned well with her role and may refuse to follow the director's instruction. All these failure modes have been highlighted above. They imply that the synthetic data obtained from observing social interactions of such alien actresses in simulated institutional settings may be a poor approximation of what observations of actual humans would deliver. The validity question looms large.

At the same time, the situation appears surprisingly far from hopeless. There are good reasons to believe that the alien actress does possess an accurate, coherent, and transferable model of human behavior that can be triggered in a DH simulation. A number of early results documented above are highly promising. The basic goal to which modern GenAIs are being trained, namely to be helpful to their human users whose intent they need to swiftly establish from an often brief and poorly worded prompt, works in our favor here. One cannot be helpful without having a reliable model of the entity they are supposed to help.

Moreover, there is much we can do to assess the DH potential thoroughly and build the necessary scaffolding to draw the most out of it. This paper sketches a research program consisting of validation efforts (back-testing, parallel testing, cross-model comparison) and infrastructure innovations (standardization, GenAI customization, benchmarking). While these are no ready-made solutions, they demonstrate the wealth of alternatives we have available to try to make the best of the DH-based synthetic data.

If we succeed, this would be nothing less than a scientific revolution. For decades, democracy research has operated under conditions of evidence scarcity. But our scarcity paradigm is not freely chosen. It is forced upon us by the high costs, slow speeds and ethical precarity of human-centric research. DH methodology, if validated, promises abundance. It could truly unlock the exploration of democracy's vast design space.

Failure, of course, is also an option. If, within the next several years, no DH study has successfully predicted performance of a novel institutional mechanism, or, even worse, if we see little progress in tackling the issues such as systematic behavioral divergence between simulated personas and their human counterparts, we should update negatively on the DH promise. However, this requires that we actually undertake the validation and standardization efforts described above. Otherwise, we cannot distinguish between 'DH methodology doesn't work' and 'true DH methodology has never been tried.'

This paper sketches the *What If World* as a thought experiment to consider the DH full potential. But neither full success nor full failure appear that likely. Probably, some limitations will remain. For instance, useful DH may remain constrained to settings with limited time horizons due to their inability to learn the way humans do. Any society-level or long-term simulations would then remain a pipe dream. However, this would not mean anything like a complete failure of the whole methodology. With stakes as high as they are in institutional research, every marginal improvement matters. Every bit of progress beyond the *status quo* is extremely valuable. The sooner the better. Also, a dynamic technology like AI may break through its constraints unexpectedly. We had better be ready at that point.

**Statement on the use of generative AI:** During the preparation of this work, the author used Claude Opus 4.5 to elicit feedback and improve his language and style. The author reviewed and edited all the content of the paper as needed, taking full responsibility for it.